\documentclass[aps,prd,twocolumn,superscriptaddress,showpacs]{revtex4}
\usepackage{graphicx}
\usepackage{dcolumn}
\usepackage{bm}
\usepackage{natbib}
\usepackage{multirow}
\usepackage{epsfig}
\usepackage{amsmath}

\newcommand{\beq}{\begin{equation}}
\newcommand{\eeq}{\end{equation}}
\newcommand{\beqa}{\begin{eqnarray}}
\newcommand{\eeqa}{\end{eqnarray}}

\newcommand{\lyaf}{Ly-$\alpha$ forest}
\newcommand{\be}{\begin{equation}}
\newcommand{\ee}{\end{equation}}
\newcommand{\ns}{n_{\rm s}}
\newcommand{\eV}{\rm eV}
\newcommand{\mpc}{\rm Mpc}

\begin{document}

\title{Cosmological parameters from combining the
Lyman-$\alpha$ forest with CMB, galaxy clustering and SN constraints}
\author{Uro\v{s} Seljak}
\affiliation{Department of Physics, Princeton University, Princeton NJ 08544, U.S.A.}
\affiliation{International Center for Theoretical Physics, Trieste, Italy}
\author{An\v{z}e Slosar} 
\affiliation{Faculty of Mathematics and Physics, University of Ljubljana, Slovenia}
\author{Patrick McDonald}
\affiliation{Canadian Institute for Theoretical Astrophysics, University of Toronto, ON M5S 3H8, Canada}

\date{\today}

\begin{abstract}
  We combine the Ly-$\alpha$ forest power spectrum (LYA) from the
  Sloan Digital Sky Survey (SDSS) and high resolution spectra with
  cosmic microwave background (CMB) including 3-year WMAP, and
  supernovae (SN) and galaxy clustering constraints to derive new
  constraints on cosmological parameters.  The existing LYA power
  spectrum analysis is supplemented by constraints on the mean flux
  decrement derived using a principle component analysis for quasar
  continua, which improves the LYA constraints on the linear power.
  We find some tension
  between the WMAP3 and LYA power spectrum amplitudes, at the $\sim 2
  \sigma$ level, which is partially alleviated by the inclusion of
  other observations: we find $\sigma_8=0.85\pm 0.02$ compared to
  $\sigma_8=0.80\pm 0.03$ without LYA. For the slope we find
  $\ns=0.965\pm0.012$.  We find no evidence for the running of the
  spectral index in the combined analysis, ${\rm dn/d\ln k}=-(1.5\pm
  1.2)\times 10^{-2}$, in agreement with inflation.  The limits on the
  sum of neutrino masses are significantly improved: $\sum
  m_{\nu}<0.17\eV$ at 95\% ($<0.32\eV$ at 99.9\%).  This result, when
  combined with atmospheric and solar neutrino mixing constraints,
  requires that the neutrino masses cannot be degenerate,
  $m_3/m_1>1.3$ (95\% c.l.).  Assuming a thermalized fourth neutrino
  we find $m_s<0.26\eV$ at 95\% c.l. and such neutrino cannot be an
  explanation for the LSND results.  In the limits of massless
  neutrinos we obtain the effective number of neutrinos  
  $N_\nu^{\rm eff}=5.3^{+0.4}_{-0.6}{}^{+2.1}_{-1.7}{}^{+3.8}_{-2.5}$
  and $N_\nu^{\rm eff}=3.04$ is allowed only at 2.4 sigma.  The constraint on the
  dark energy equation of state is $w=-1.04\pm 0.06$.  The constraint
  on curvature is $\Omega_{\rm k}=-0.003\pm 0.006$.  Cosmic strings
  limits are $G\mu <2.3\times 10^{-7}$ at 95\% c.l. and correlated
  isocurvature models are also tightly constrained.

\end{abstract}


\maketitle

\setcounter{footnote}{0}

\section{Introduction}

The latest WMAP results of cosmic microwave background (CMB) temperature
and polarization anisotropies from 3 years of observations have
presented a remarkably clear view of the universe in its early epoch
around redshift 1100
\cite{2006astro.ph..3451H,2006astro.ph..3450P}.  These data, combined
with other tracers such as galaxy clustering and supernovae, can be
satisfied within a simple cosmological model in which the universe is
filled with cold dark matter and dark energy and the initial
conditions are generated through a process like inflation
\cite{2006astro.ph..3449S}.  One of the main new results from this
analysis relative to the 1st year WMAP analysis \cite{2003ApJ...583....1B}
is the reduced value of the optical depth due to reionization, which
follows from a more sophisticated modeling and better statistics of
CMB polarization data.  Many other parameters are affected by this
change, such as a reduced value of the amplitude of fluctuations
$\sigma_8$ and increased evidence for a red spectral index of
primordial fluctuations, $\ns<1$ (note, however, that the optical depth 
constraint accounts for only part of the change in parameters).

To increase the statistical power of constraints the WMAP team added
other tracers, such as small scale CMB
\cite{2004ApJ...609..498R,2004MNRAS.353..732D,2002AAS...20114004K,2005astro.ph..7503M},
galaxy clustering from 2dF and SDSS
\cite{2005MNRAS.362..505C,2004ApJ...606..702T,2005ApJ...633..560E},
supernovae \cite{2004ApJ...607..665R,2006A&A...447...31A} and weak
lensing \cite{2005A&A...429...75V,2005PhRvD..71d3511S}.  All of the
tracers used in their analysis trace large scale structure on scales
above ten megaparsecs. However, there are many physically motivated
models that can only be distinguished using small scale information.
One is the running of the spectral index, $\alpha={\rm dn/d\ln k}$.
Inflation generically predicts that there should be no running at the
observable level in the current data.  Therefore, any detection of
running would be a major surprise and would force us to either accept
models of inflation with an unusual shape of the potential with large
third derivative or abandon it in favor of a better theory. In the
latest WMAP analysis the data are marginally better fitted with
non-zero running, $\alpha=-0.06\pm 0.03$, but this analysis did not
include small scale information, so the error is large.

Another class of models where small scale clustering information helps
are those with massive neutrinos.  These suppress power in dark matter
clustering on small scales because of their free streaming, which
erases their own fluctuations on scales below the free streaming
length.  This in turn slows down the growth of cold dark matter
structure on the same scales, leaving an imprint in the dark matter
power spectrum.  For the relevant range of neutrino masses below 1eV
the full extent of the suppression happens on megaparsec scales.
From atmospheric and solar neutrino mixing detections we know that at
least one neutrino has mass above 0.05eV and another above 0.01eV 
\cite{Ashie:2005ik,Maltoni:2004ei},
 but these experiments only
tell us the mass difference squared and cannot provide the absolute
scale for neutrino mass. Cosmological observations weigh neutrinos and
can provide an absolute scale for their mass.  Thus cosmological
observations can answer one of the most important issues in neutrino
physics today, the nature of the  mass hierarchy of neutrinos.

It is clear that to improve upon these constraints one should
determine the fluctuation amplitude on smaller scales than probed by
CMB, galaxies or weak lensing.  Nonlinear evolution prevents one from
obtaining useful information at $z=0$, so one must look for probes at
higher redshift.  Of the current cosmological probes, the Ly-$\alpha$
forest (LYA) -- the absorption observed in quasar spectra by neutral
hydrogen in the intergalactic medium (hereafter IGM) -- has the
potential to give the most precise information on small scales
\cite{1998ApJ...495...44C}.  It probes fluctuations down to megaparsec
scales at redshifts 2--4, so nonlinear evolution, while not
negligible, has not erased the primordial information.

Currently the most precise measurement of the LYA power spectrum comes from
the analysis of more than 3000 spectra from the Sloan Digital
Sky Survey \cite{2006ApJS..163...80M,2005ApJ...635..761M}. 
This data set is almost two orders of
magnitude larger than available previously and has dramatically
reduced the errors on the amplitude and slope of the power spectrum
around wave-vector $k \sim 1 h \mpc^{-1}$.  
Combining this and a small amount
of higher resolution \lyaf\ data 
\cite{2000ApJ...543....1M,2005ApJ...635..761M} with other
probes has already been demonstrated to significantly improve the
constraints on the running and neutrino masses
\cite{2005PhRvD..71j3515S}.  For the running the current constraint
from WMAP 1st year and LYA is $\alpha=-0.003\pm 0.010$, which is
nearly a factor of 3 improvement over the WMAP 3 year analysis without
LYA. For neutrino mass the constraints including LYA are $\sum
m_{\nu}<0.3-0.4{\rm eV}$ at 95\% confidence level
\cite{2005PhRvD..71j3515S,2006astro.ph..2155G} in the most restrictive
6 parameter analysis.

The previous joint LYA and WMAP first year analysis in
\cite{2005PhRvD..71j3515S} suffered a significant increase in the
errors on the amplitude and slope of the power spectrum from the
\lyaf\ due to projecting out all of the poorly known nuisance
parameters such as the temperature-density relation, flux of UV
radiation and its fluctuations, filtering length, effects of galactic
winds etc.  The largest part of this increase is due to a partial
degeneracy between the mean flux decrement determined by the intensity
of the UV background and the amplitude of the power spectrum
\cite{1998ApJ...495...44C,2005ApJ...635..761M}.  While this degeneracy
is partially broken by the flux power spectrum information itself, the
error on the amplitude increases significantly over the case where the
mean flux decrement was known.

One can improve upon this by providing external constraints on the
redshift evolution of the mean flux decrement. In principle, if one
knew the shape of the quasar spectrum, one would extract the decrement
simply by comparing the actual flux level to the predicted flux level
in the absence of absorption. One way to determine the continuum level
of quasars is to identify regions with no absorption, but this is
difficult at redshifts above 2.5, where the amount of absorption is
significant and such regions are rare. Another way is to assume quasar
spectra all have the same shape and the continuum is known from
observations at lower redshifts where absorption is negligible.  In
practice the quasars do not all have the same shape, so this approach
has to be generalized.  One can use the fact that the quasar continuum
follows well defined variations and, identifying these at rest wavelengths
longer than 1216\AA, where there is no absorption, predict the shape in the 
\lyaf.  A formalization of this procedure, including the errors,
can be done by applying the principal component analysis (PCA) on the
measured spectra, possibly including those measured at low redshifts where
there is little Ly-$\alpha$ forest observed \citep{2005ApJ...618..592S}.  
This procedure leads to
a better determination of the mean flux decrement as a function of
redshift up to an overall amplitude, which cannot be determined with
this method.   
However, at low redshifts the level of absorption is
sufficiently low that the continuum can be determined simply by
connecting regions of no absorption, especially when using high
resolution spectra \cite{2000ApJ...543....1M}.  
This method therefore
provides external information on the mean flux decrement, which can be
used in the joint analysis.

The procedure described above has been applied to SDSS data and
results presented in \cite{2005ApJ...635..761M}. They show a
remarkable agreement between the PCA and power spectrum based
determination of the mean flux decrement as a function of redshift.
This agreement provides additional confidence in the results obtained
from the power spectrum analysis.  Since the PCA method gives somewhat
smaller errors on the mean flux decrement one can use these as
external constraints to better determine the amplitude of the power
spectrum.  In the current analysis the improvement is modest, 20-30\%
on both amplitude and slope, because we do not have a tight constraint on 
the overall normalization of the mean absorption.
The purpose of this paper is to apply
these LYA constraints to the joint cosmological analysis with other
data sets, including the new WMAP 3 year data. An independent analysis 
of similar type has recently been performed by \cite{Viel:2006yh}. 

\section{Analysis} 

We combine the constraints from the Ly-$\alpha$ forest (LYA)
\cite{2005ApJ...635..761M} with the SDSS galaxy clustering analysis
(SDSSgal) \cite{2004ApJ...606..702T}, SDSS luminous red galaxy
constraints on the acoustic peak (SDSS BAO) \cite{2005ApJ...633..560E},
2dF galaxy clustering analysis (2dF) \cite{2005MNRAS.362..505C}, Gold
sample of supernovae from \cite{2004ApJ...607..665R} and SNLS
supernovae sample \cite{2006A&A...447...31A}, CMB power spectrum
temperature and polarization observations from WMAP 3 year analysis
(WMAP3) \cite{2006astro.ph..3451H,2006astro.ph..3450P} and from
smaller scale experiments, including Boomerang-2k2
\cite{2005astro.ph..7503M}, CBI \cite{2004ApJ...609..498R}, VSA
\cite{2004MNRAS.353..732D}, and ACBAR \cite{2002AAS...20114004K}
(CMBsmall).  Contrary to the WMAP approach we do not include the SDSS
bias constraints \cite{2004astro.ph..6594S} nor weak lensing
constraints \cite{2005A&A...429...75V}, because their statistical
power is weak when the LYA power spectrum is included and we prefer
not to cloud the interpretation of the results.  We mention however
that both of these constraints prefer higher normalization of
$\sigma_8$ than the recent WMAP3 results and provide additional
support for the results found in this paper. Moreover, for the case of
weak lensing, the recent detection of intrinsic alignment
\cite{2006MNRAS.367..611M} may push the weak lensing normalization
even further up from existing numbers, which do not account for this
effect.

Our most general cosmological parameter space is 
\begin{multline}
p=(\omega_{\rm b},\omega_{\rm dm},\Omega_{\rm k}, \theta, \tau,
f_{\nu}, N_\nu^{\rm eff}, {\rm w}, A, \ns, \alpha,r, \\A_{\rm  iso},  G\mu),
\end{multline}
where $\omega_{\rm b}=\Omega_{\rm b}h^2$, with $\Omega_{\rm b}$ baryon
density in units of the critical density, $\omega_{\rm dm}=\Omega_{\rm
  dm}h^2$ with $\Omega_{\rm dm}$ dark matter density in units of the
critical density, $h$ is the Hubble's constant in units of 100 km
s$^{-1} \mpc^{-1}$, $\Omega_{\rm k}=1-\Omega_{\rm tot}$, where
$\Omega_{\rm tot}$ is the total density of the universe in units of
the critical density, $\theta$ is the angular scale of the acoustic
horizon, $\tau$ is the optical depth to the surface of last
scattering, $f_\nu$ is the fraction of dark matter in neutrinos (the
rest is assumed to be cold dark matter), $N_\nu^{\rm eff}$ is the the
energy density in relativistic background expressed in units of
massless neutrino families and w the equation of state of dark energy,
(which in general can be time dependent, but in this paper we only
explore it as a constant).  Parameters $A_{\rm iso}$ and $G\mu$
describe the iso-curvature modes and string contribution (see section
\ref{isocos}). The power spectrum of primordial fluctuations is
assumed to be of the form
\begin{equation}
    \nonumber
    P_i(k) = A \left(\frac{k}{k_0}\right)^{\ns + \alpha \log (k/k_0)/2},
\end{equation}
with the pivot point chosen to be $k_{0}=0.05/$Mpc.  We use two
additional amplitude parameterizations, which are derived parameters:
$\sigma_8$ is the rms of fluctuations in the linear density field
smoothed at 8$h^{-1}{\rm Mpc}$ scale at $z=0$, and 
$\Delta^2(k=0.009{\rm s/km},z=3)$
describing the fluctuations at roughly one megaparsec scale at $z=3$.
Finally, $r$ is the ratio of
tensor to scalar power spectrum amplitude at pivot point $k_0$. 

We begin the exploration with the simplest 6 parameters $(p=
\omega_{\rm b},\omega_{\rm dm}, \theta, \tau, A, \ns)$
and then proceed by adding additional parameters, one at a time.  We
have tested the analysis using two independent Monte Carlo Markov
Chain (MCMC) codes.  The first code is described in the previous analysis
of the Ly-$\alpha$ forest \cite{2005PhRvD..71j3515S} and uses CMBFAST
\cite{1996ApJ...469..437S}.  The second is the standard CosmoMC package
\cite{2002PhRvD..66j3511L} using CAMB \cite{2000ApJ...538..473L}. We
find a very good agreement between the two codes.  We also compare our
results whenever possible to the WMAP analysis, finding very good
agreement in the analysis of WMAP data.  We find some small
discrepancies when other data sets are combined with WMAP. For
example, in our WMAP+SDSSgal analysis we find $\sigma_8=0.803$ compared
to 0.772 in the WMAP paper \cite{2006astro.ph..3449S}, which we suspect
can be explained by the different treatment of nonlinear effects in
the SDSSgal data, as discussed below.

For galaxy surveys we adopt a somewhat more conservative approach in 
the modeling of nonlinear effects than 
what was done previously. The nonlinear biasing effects are 
argued to be well described by the expression given in \cite{2005MNRAS.362..505C},
\begin{equation}
P_{\rm gal}(k)=b^2 { 1+Qk^2 \over 1+A_gk} P_{\rm lin, dm}(k).
\end{equation}
We use this expression with $A_g=1.4$ and marginalize over the bias
parameter $b$, but rather than adopting a fixed value for $Q$ we treat
it as a free parameter. For the 2dF power spectrum we adopt a Gaussian
prior $Q=4.6\pm 1.5$, as found in \cite{2005MNRAS.362..505C}.  For
the SDSS galaxy power spectrum we use the same approach with $Q=10\pm 5$
and $Q>0$. We use a higher mean value for $Q$ because the SDSS data
were analyzed in real space, where nonlinear effects are stronger and
the value $Q=10$ fits well the standard nonlinear power spectrum from
N-body simulations (at a percent level).  Exploring the
allowed range of $Q$ in numerical simulations we found that even
values above 20 are possible in real space.  Therefore, we argue that
using a fixed value for $Q$ is not justified given the uncertainties
in the nonlinear galaxy biasing.  Instead we allow for a wider range
of $Q$ to account for the uncertainty and let the data choose the best
value.  In both cases we use the data up to $k=0.15h/{\rm Mpc}$, so the
nonlinear effects are small, but not negligible.

For baryonic oscillations (BAO) we use the recent measurements of the
baryonic peak in the Luminous Red Galaxies sample from the SDSS as
published in \citep{2005ApJ...633..560E}. To compare theoretical power
spectra with the measured correlation function, we follow the
prescription of the authors of the data: first we interpolate between
the linear power spectrum and the ``no-wiggle'' equivalent using the
weighting function $\exp(-(ak)^2)$ with $a=7 h^{-1} {\rm Mpc}$. For
the ``no-wiggle'' spectrum we took either the approximation from
\citep{1998ApJ...496..605E} (as used in \citep{2005ApJ...633..560E})
or the convolution with suitable top-hat in the $\log P(k)$-$\log k$
space. This is because the approximation of
\citep{1998ApJ...496..605E} is not available for all cosmological
models.  We also compared the results to those found when this step is
ignored.  In general we find that the effects from these different
treatments on the final results are below one sigma, but when this is
not the case we explicitly discuss it.  Then the non-linear
corrections are applied using the HaloFit \cite{2003MNRAS.341.1311S}
package. The resulting power spectrum is converted to the
corresponding correlation function. Finally, the red-shift space
distortion and non-linear bias effects are accounted for by
multiplying the correlation function $\xi(r)$ by $(1+0.06/(1+(0.06
h/{\rm Mpc}\times r)^6))^2$, in agreement with $N$-body simulations.
This correlation function is then compared with the measurement using
the full covariance between errors. The amplitude of the correlation
function is marginalized over.

\section{Results}

\subsection{Basic cosmological parameters}

Table \ref{tab0} shows the results of the analysis with and without
LYA for various parameter combinations, starting with the
simplest 6 parameter model
$p=(\tau,\omega_{\rm b},\omega_{\rm dm},\Omega_{\lambda}, A,
\ns)$.  For most parameters adding LYA reinforces the conclusions
derived by WMAP.  However, LYA prefers a high value of the
normalization and as a consequence other parameters are also affected.
For example, the preferred optical depth in the joint analysis
increases to $\tau=0.11$ and the normalization increases to
$\sigma_8=0.85$.
%In Figure \ref{twod} we show the most correlated combination of
%parameters present in the simplest 6-parameter model.

\begin{table}
\begin{tabular}{ccc}
parameter & ALL DATA & ALL DATA - LYA \\
\hline
&&\\
$\omega_{\rm b}   $&$ 0.0230^{+0.0006}_{-0.0006}{}^{+0.0011}_{-0.0011}{}^{+0.0017}_{-0.0019} $&$0.0224^{+0.0007}_{-0.0006}{}^{+0.0012}_{-0.0013}{}^{+0.0018}_{-0.0019}$ \\
&&\\
$\omega_{\rm dm}  $&$ 0.117^{+0.003}_{-0.002}{}^{+0.005}_{-0.005}{}^{+0.007}_{-0.008} $&$0.114^{+0.003}_{-0.003}{}^{+0.006}_{-0.005}{}^{+0.009}_{-0.008}$   \\
&&\\
$h                $&$ 0.705^{+0.013}_{-0.013}{}^{+0.025}_{-0.023}{}^{+0.038}_{-0.038} $&$0.703^{+0.013}_{-0.013}{}^{+0.025}_{-0.028}{}^{+0.037}_{-0.037}$ \\
&&\\
$\tau             $&$ 0.108^{+0.010}_{-0.010}{}^{+0.039}_{-0.043}{}^{+0.063}_{-0.069} $&$0.077^{+0.014}_{-0.015}{}^{+0.045}_{-0.046}{}^{+0.083}_{-0.064}$ \\
&&\\
$\ns              $&$ 0.964^{+0.012}_{-0.012}{}^{+0.025}_{-0.026}{}^{+0.037}_{-0.038} $&$0.951^{+0.013}_{-0.013}{}^{+0.027}_{-0.027}{}^{+0.041}_{-0.041}$ \\
&&\\
$\sigma_8         $&$ 0.847^{+0.022}_{-0.022}{}^{+0.042}_{-0.045}{}^{+0.070}_{-0.062} $&$0.798^{+0.030}_{-0.030}{}^{+0.059}_{-0.053}{}^{+0.083}_{-0.075} $ \\
\hline
&&\\
$r$ & $<0.22\ (<0.37)$ & \\
&&\\
$\alpha$ &  $-0.015^{+0.012}_{-0.012}{}^{+0.023}_{-0.021}{}^{+0.036}_{-0.029}$ & \\
&&\\
$N_\nu^{\rm eff}$ & $5.3^{+0.4}_{-0.6}{}^{+2.1}_{-1.7}{}^{+3.8}_{-2.5}$  & \\
&&\\
$\sum m_{\nu}$ & $<0.17\eV\,(<0.32\eV)$ & \\
&&\\
$m_{s}$ & $<0.26\eV\,(<0.43\eV)$ & \\
&&\\
$\Omega_{\rm k}$ & $-0.003^{+0.0060}_{-0.0061}{}^{+0.0109}_{-0.0122}{}^{+0.0157}_{-0.0180}$ & \\
&&\\
w & $-1.040^{+0.063}_{-0.063}{}^{+0.124}_{-0.130}{}^{+0.178}_{-0.208}$ &  \\ 
&&\\
$A_{\rm iso,bar}$& $-0.06^{+0.18}_{-0.18}{}^{+0.35}_{-0.34}{}^{+0.50}_{-0.55}$ &\\
&&\\
$A_{\rm iso,CDM}$ & $-0.007^{+0.034}_{-0.035}{}^{+0.068}_{-0.067}{}^{+0.110}_{-0.104}$ \\
&&\\
$G\mu$ & $<2.3\times 10^{-7} (<2.9\times 10^{-7})$ &\\
\end{tabular}
\caption{Parameter constraints for the simplest 6-parameter
  model. Parameters below the horizontal line ($r$\ldots$G\mu$) were
  constrained individually, varying the basic 6 parameters and one
  extra parameter at a time. Limits are 1, 2 and 3 sigma; when only upper
  limits are given, these are at 95\% and 99.9\%. }
\label{tab0}
\end{table}

%\begin{figure}
%\includegraphics[width=\linewidth]{wmap3allo_2D.ps}

%\caption{This figure shows the most correlated parameters left in the
%  standard 6-parameter model}

%\label{twod}
%\end{figure}

\subsection{Amplitude of fluctuations: are the data sets compatible?}

The parameter where there is the most tension between the analyses with and
without the Ly-$\alpha$ forest is the amplitude of fluctuations. In terms of
$\sigma_8$, in \cite{2005PhRvD..71j3515S} this parameter was found to
be $\sigma_8=0.90\pm 0.03$, but this value was driven by the WMAP
optical depth prior, as seen in Figure 1 of that paper, which suggests
that for the new value $\tau=0.09$ one would find $\sigma_8=0.84$. In
our new analysis we find $\sigma_8=0.85\pm 0.02$, which is in a good
agreement with this expectation.  This should be compared to
$\sigma_8=0.75\pm 0.06$ for WMAP alone and $\sigma_8=0.80\pm 0.03$
for WMAP combined with everything else in our analysis. There is some tension
between the WMAP and LYA values and the most relevant question is
whether this can be explained as a statistical fluctuation or as a
signal of a systematic problem in the data.

\begin{figure}
\begin{center}
 \epsfig{file=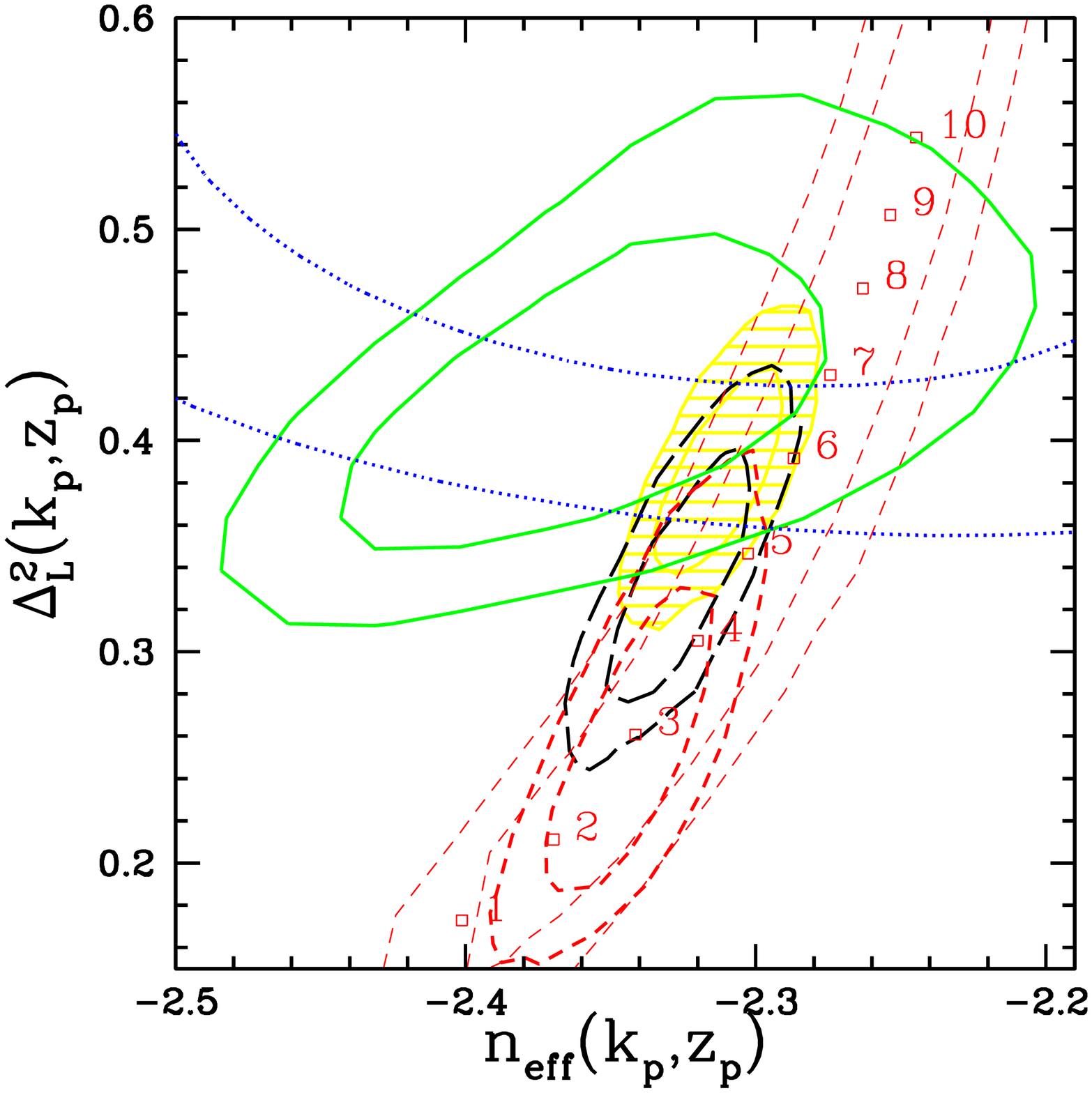,width=\linewidth} \\
 \epsfig{file=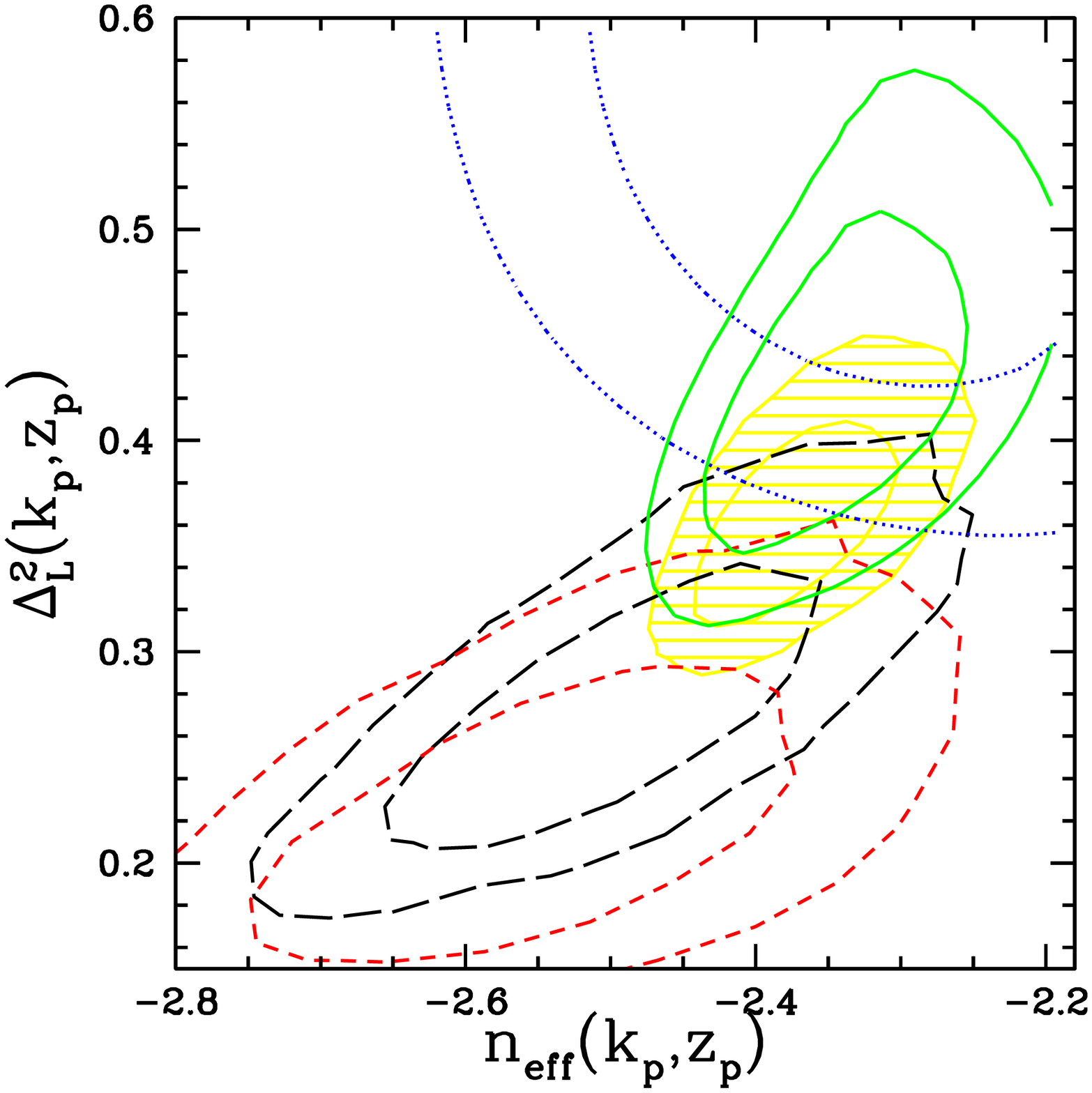,width=\linewidth}
\end{center}
\caption{Constraints on effective slope $n_{\rm eff}$ and
amplitude $\Delta^2$ of the linear power at $k_p=0.009~{\rm s/km}$, $z=3$. 
The top panel corresponds to models with power-law primordial
fluctuations, while the bottom includes running of the spectral index.
In all cases 68\% and 95\% contours are shown.  Red, short-dashed contours
are for WMAP3, black, long-dashed for WMAP3+SDSSgal+
SN, green, solid for LYA only, and yellow, solid contours with
horizontal stripes are for all of these constraints combined.
The blue, dotted contours show an alternative
\lyaf\ analysis using only high resolution spectra \cite{2004MNRAS.354..684V}. 
Top panel also shows the WMAP3 constraints when $N_{\nu}^{\rm eff}$ is allowed to be 
free (thin red short-dashed) versus being fixed at $N_{\nu}^{\rm eff}=3.04$ (thick 
red short-dashed), as well as the central values of WMAP3 contours
for different values of $N_{\nu}^{\rm eff}$ from 1 to 10. 
}
\label{fig1}
\end{figure}

The \lyaf\ effectively constrains only the amplitude and slope of the
linear power spectrum at pivot point $k=0.009\ {\rm s\ km}^{-1}$ and
$z=3$.  Projections to other scales or redshifts are sensitive to
cosmological parameters that are essentially unconstrained by the
\lyaf.  On the other hand, WMAP alone can make a well-controlled
projection into the LYA amplitude-slope plane, so we make our
comparison in this plane.  Constraints from various experiments are
shown in Figure \ref{fig1}.  The top panel is for the case of no
running using the basic 6 parameters.  In this case WMAP data alone
already provide a strong constraint in this plane. We see that LYA
data prefer a higher value for the amplitude than WMAP and about the
same value for the slope, but the 2-sigma contours have a considerable
overlap.  The agreement is improved if other probes, such as SDSS
galaxy clustering, are added to WMAP data, which is also shown in the
figure.  In terms of differences in $\chi^2$ between fitting for
CMB+SDSSgal+SN and LYA separately versus jointly we find that a joint
fit increases the total $\chi^2$ by 3.8, also suggesting less than
2-sigma discrepancy.  Finally, we see that the \lyaf\ constraints
using only high resolution spectra
\cite{2003astro.ph..8103K,2002ApJ...581...20C}, while considerably
weaker compared to our analysis, also prefer a higher normalization.
Here we have taken the values given in \cite{2004MNRAS.354..684V},
including the 15\% error on the amplitude, marginalized separately in
two redshift bins. We note that our own analysis of these data gives a
somewhat smaller error on the amplitude and a somewhat larger error on
the slope, but the basic conclusions remain the same.

The situation is similar when running is added as a parameter.
One can see that WMAP3+SDSSgal+SN errors are comparable to LYA.  In
this case the WMAP3 data prefer large negative running, so even though
their errors significantly increase in this plane, the mean value has
been pulled towards lower values for both amplitude and slope and the
agreement with LYA is actually slightly worse than before.
Still, there is considerable overlap of 2-sigma contours.  The
difference in $\chi^2$ between fitting WMAP3+SDSSgal+SN and LYA separately
and jointly is now 6.

So is there any evidence for inconsistency between WMAP3 and LYA data?
From the comparisons made above it would seem that in the simplest
6-parameter models the discrepancy is around 2-sigma.  However, it is
important to recognize that any attempt to answer this is an
a-posteriori attempt and can only provide qualitative answers.  For
example, if we focus on a single parameter like amplitude, which is
chosen a posteriori from N parameters on the basis of the fact that it
shows the most discrepancy, then we will obtain biased results: 
comparing two experiments in N parameters and selecting the most
discrepant one on the basis of one out of p criteria (for example, if
one can choose from larger than and smaller than then p=2) will on
average give one parameter in the 1/pN probability corner.  In our case
LYA most naturally measures amplitude and slope at one megaparsec, so
N=2 and p=2 and the statistical significance of discrepancy is further
diluted by a factor of 4.   
Our conclusion is thus that
there is no compelling evidence that the differences between LYA and WMAP3
cannot be explained as a normal statistical fluctuation and we proceed under
this assumption.  We return to this in the discussions.

\subsection{Testing inflationary models: spectral index and tensors}

LYA prefers a higher amplitude at a megaparsec scale and this
also pushes the primordial slope upwards.  The effect is small, but
relevant for the question of whether scale invariant spectrum with
$\ns=1$ is statistically excluded.  We find $\ns=0.965\pm 0.012$ versus 
$\ns=0.951\pm 0.013$ without LYA, so about one sigma change in the 
value. A very similar value was obtained in \cite{Sanchez:2005pi}.

Adding tensors to the analysis does not improve the fit. We find
\begin{equation}
r<0.22\, {\rm at}\,  95\% {\rm c.l.} \, (<0.37\, {\rm at} \, 99.9\%),
\end{equation}
slightly stronger than $r<0.28-0.3$ at 95\% c.l. from WMAP3+SDSSgal or
WMAP3+2dF analysis \cite{2006astro.ph..3449S}.

\begin{figure}
%% 2Dhisto.py -o nsr.eps/PS -c 14 -d 10 -y "n\\ds\\u" -x "r" -C -h 0.3 -w 1 -n 29 -m 29 tens.txt 
\begin{center}
\includegraphics[width=\linewidth]{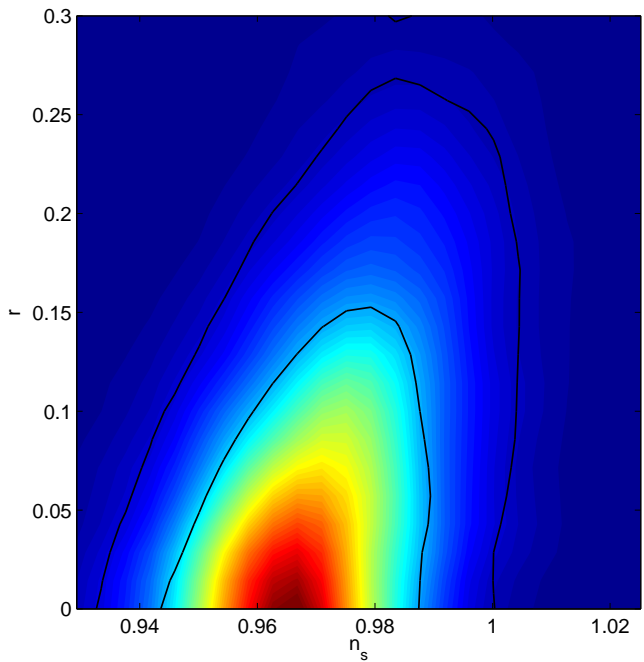}
\end{center}
\caption{This figure shows constraints on the $n_s$--$r$
  plane. Contours enclose 68\% and 95\% probability. }
\label{fig:nsr}
\end{figure}

In Figure \ref{fig:nsr} we show the constraints in $(\ns, r)$ plane. This 
plane is of interest for inflation model builders, since many of 
the inflationary models can be classified according to their
predictions in this plane. We see that models with $\ns>1$ are 
disfavored at $r=0$, but there is a positive correlation between 
$r$ and $\ns$. 

\subsection{Testing inflationary models: running of the spectral index}

The issue of the running of the primordial slope continues to be
controversial. The latest WMAP analysis finds the strongest
evidence for running so far, particularly when combined with other
data sets such as small scale CMB or SDSSgal, e.g.  %%$dn /d\ln k
$\alpha=-0.066^{+0.026}_{-0.032}$ for WMAP+CBI+VSA.  The dynamic range of
scales is limited by CMB and galaxy clustering, and the corresponding
error is large.  It was previously shown that the error is
significantly reduced when SDSS LYA and WMAP first year data were 
combined, giving %%$dn%%/d\ln k
$\alpha=-0.003\pm 0.010$ \cite{2005PhRvD..71j3515S}.

Here we redo the analysis using the new
WMAP3 data and our LYA analysis. Together with all the other data 
we find
\begin{equation}
\alpha=-0.015^{+0.012}_{-0.012}{}^{+0.023}_{-0.021}{}^{+0.036}_{-0.029}
\end{equation}
The combined analysis therefore continues to show no evidence for
running, in agreement with theoretical expectations from inflation.
The errors are improved relative to the WMAP 3 year analysis without
LYA, but have not improved relative to LYA + first year WMAP, which
also did not show any evidence for running.  At this point, a
conservative conclusion
is that the results of the WMAP team analysis without LYA are a
statistical fluctuation which is eliminated by adding 
the LYA data, which are
much more constraining than any combination without LYA.  We note that
2-sigma deviations from the true value on one of a dozen parameters are bound
to happen in almost any analysis and the above discussion
of dangers of a-posteriori assessment of statistical significance also
applies here.

\subsection{Number of relativistic neutrino families}

In the standard model the main relativistic components of energy
density are photons and neutrinos. Neutrinos decouple earlier than
photons, which receive additional entropy transfer from
electron-positron annihilation, and as a result photons are hotter
than neutrinos, $T_{\nu}=(4/11)^{1/3}T_{\gamma}$.  The neutrino
contribution to the energy density is usually described in terms of
an effective number of neutrino families assuming the temperature
relation above. The standard model predicts $N_\nu^{\rm eff}=3.04$ instead of
3 because some of the entropy transfer goes to neutrinos that are not
completely decoupled by the time of electron-positron annihilation.
More generally, we use $N_\nu^{\rm eff}$ as a parametrisation of relativistic
energy density in the early universe, although it should be 
noted that effects of neutrinos may differ from other relativistic 
particles \cite{2004PhRvD..69h3002B}.  If we assume $N_\nu^{\rm eff}$ neutrino
families then the analysis finds
\begin{equation}
N_\nu^{\rm eff} = 5.3^{+0.4}_{-0.6}{}^{+2.1}_{-1.7}{}^{+3.8}_{-2.5}
\label{eq:nnu}
\end{equation}

We find that $P(N_\nu^{\rm eff})<3.04$ is $8.5\times10^{-3}$.  If only integer values
of the number of neutrinos are allowed, we find that $3:4:5$ neutrinos
are allowed with relative probabilities $1:\sim 6.4:\sim 9.7$.
How robust is the evidence against 3 neutrino families? 
Adding the HST
constraint \cite{2001ApJ...553...47F} in the form of a Gaussian prior
on Hubble's constant $h=0.72\pm0.08$ lowers $N_\nu^{\rm eff}$ to
$4.8^{+0.4}_{-0.5}{}^{+1.6}_{-1.4}{}^{+3.0}_{-2.1}$ (but leaves the
lower 3-sigma bound largely unchanged). This is because there is a
strong degeneracy between the number of relativistic degrees of
freedom and the dark matter energy density, as any additional
relativistic component can be compensated by additional dark matter in
order to keep matter-radiation equality unchanged
\cite{2005astro.ph.11500D} and so strong degeneracy exists with $h$
and $\omega_{\rm dm}$.  For similar reasons, the constraint is also
partially sensitive to the BAO treatment.  Without BAO we find that
$N_\nu^{\rm eff}=6.0^{+1.4}_{-1.4}{}^{+2.9}_{-2.4}{}^{+3.8}_{-3.3}$.

Removing Lyman
alpha data lowers the number to
$3.9^{+0.4}_{-0.6}{}^{+2.1}_{-1.7}{}^{+4.7}_{-2.7}$, so the 
evidence is significantly weakened.  
However, this is with both SDSS and 2dF galaxy power spectra 
and there is significant variation between the two. 
We find that the WMAP3 data together with supernovae data and
SDSS main sample galaxy data favour high
$N\nu=7.8^{+1.1}_{-0.7}{}^{+2.3}_{-3.2}{}^{+2.9}_{-5.4}$, while
replacing SDSS galaxy data with 2dF data favours much lower number of
neutrinos: $N_\nu^{\rm
  eff}=3.2^{+0.5}_{-1.0}{}^{+3.6}_{-2.3}{}^{+6.4}_{-3.0}$ (note that
we cannot reproduce the results in Table 10 of
\cite{2006astro.ph..3449S}, however, the implausible asymmetry of
their errors suggests some problem with those results).  
Excluding both the 2dF and SDSS main galaxy samples from the mix of
all data results in $N_\nu^{\rm
  eff}=5.2^{+0.5}_{-0.6}{}^{+2.1}_{-1.8}{}^{+3.9}_{-2.6}$, which is
only a minor change compared to all datasets.

We can conclude that the unexpectedly large result for the number of
neutrino families does not rely on a single experiment as several
combinations of data favour high values of relativistic energy
density, albeit with an expected lower statistical significance. 
However, at the moment the evidence is strongest with the Lyman alpha
data. 
Top panel of figure 
\ref{fig1} shows why relaxing $N_\nu^{\rm eff}$ improves the 
fit. When the parameter increases it leads to a higher amplitude and slope 
in LYA plane relative to the standard value, 
reducing the discrepancy between LYA and WMAP3 data. 
This is the only parameter we have found that has this effect. 

Our
result of combining all datasets disfavors just three mass-less
neutrinos, favouring higher numbers of neutrinos instead.  Big Bang
Nucleosynthesis (BBN) in general tends to favour a lower number of
neutrinos.  In particular, \cite{Cyburt:2004yc} obtain $N_\nu^{\rm
  eff}=3.14^{+0.70}_{-0.65}$ from BBN data alone and $N_\nu^{\rm
  eff}=3.24^{+0.61}_{-0.57}$ by using the CMB photon-to-baryon ratio,
while \cite{Olive:1998vj} get limits from $N_\nu^{\rm eff}<3.3$ at
95\% c.l. to $N_\nu^{\rm eff}<6$ and 95\% c.l. depending on the
details of the analysis. At the moment $N_\nu^{\rm eff}=4$ is the
integer number of neutrinos most compatible with the data. Higher
$N_\nu^{\rm eff}$ is at some tension with BBN constraints, while lower
$N_\nu^{\rm eff}$ has trouble with the datasets discussed in this
paper.  However, newly discovered systematic issues in various
datasets might still shift values around. Finally, the possibility of
a statistical fluke should not be discounted, as the 
standard value is within 2.4 sigma of the measured value. 

%% These distributions are shown in Figure \ref{fig:nbao}.  We conclude
%% that the constraints given in equation \ref{eq:nnu} should be viewed
%% as preliminary and more work on the details of nonlinear evolution is
%% needed to confirm these results.

%% \begin{figure}
%%   \begin{center}
%%     \includegraphics[width=\linewidth]{numneutX_2D.ps} \\
%% %%    \includegraphics[width=\linewidth]{numneutX.ps}
%%   \end{center}
%%   \caption{This figure shows constraints on the
%%     $\omega_{\rm dm}$--$N_{\nu}$ plane for ALL data used (thick solid) line,
%%     the same with alternative ``no-wiggle'' spectrum for BAO treatment
%%     (thin solid) (see text) and all datasets except BAO data
%%     (dotted).  }
%%   \label{fig:nbao}
%% \end{figure}

\subsection{Massive neutrinos}

As mentioned in the introduction, small scale clustering combined with
large scales and CMB is a very efficient way to weigh neutrinos, since
these suppress the growth of structure in CDM.  One can expect that
adding LYA leads to a significant improvement in the constraints.
Since the Ly-$\alpha$ forest prefers a high normalization compared to the CMB 
and the two
data sets are barely compatible with each other for massless
neutrinos, adding massive neutrinos which suppresses power on the LYA scale 
would further increase the discrepancy and
make them statistically incompatible, thus providing a tight limit on
the neutrino mass.  We find
\begin{equation}
\sum m_{\nu}<0.17\eV\, {\rm at}\,  95\% {\rm c.l.} \, (<0.32\eV \, {\rm at} \, 99.9\%).
\end{equation}
These results depend somewhat on the treatment of the BAO data. If
smoothing of the spectrum is used the limits raise to $0.35\eV$ at
99.9\% c.l. and to $0.40\eV$ if BAO data are not included at all.
Either way, this result appears to rule out the claimed
Heidelberg-Moscow experiment of neutrino-less double beta decay
\cite{Klapdor-Kleingrothaus:2004wj}.

\begin{figure}
  \centering
  \includegraphics[angle=-90,width=\linewidth]{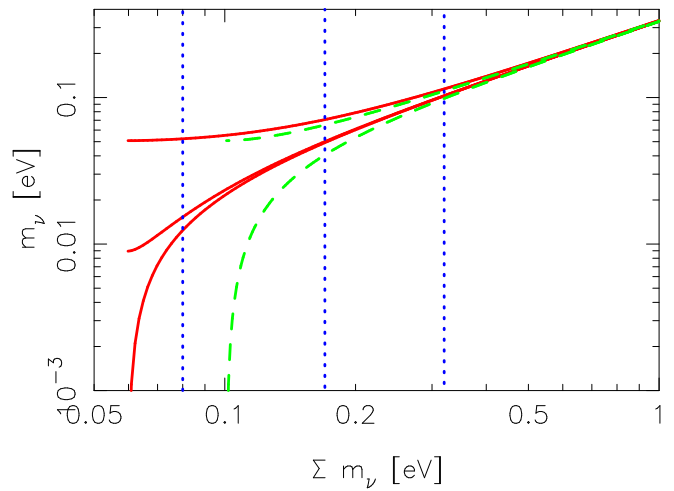}
  \caption{This figure shows the masses of individual neutrinos 
    plotted against the sum of masses which is constrained by the
    cosmological data. Solid line (red) is for normal hierarchy, while dashed
    (green) is for the inverted hierarchy (note that two upper to mass
    eigenstates are nearly degenerate in this case). Vertical dotted lines 
correspond to
    our 68\%, 95\% and 99.9\% confidence limits. The assumed mass squared
    differences were taken to be $\Delta m_{12}^2=8\times
    10^{-5}{\rm eV}^2$ and  $\Delta m_{23}^2=2.5 \times 10^{-3}{\rm eV}^2$. }
  \label{fig:bongo}
\end{figure}

If we combine the constraint on the sum of the masses with the
existing neutrino mixing constraints, which are $\Delta m^2_{12}=8\times
10^{-5}{\rm eV}^2$ for solar and $\Delta 
m^2_{13}=2.5\times 10^{-3}{\rm eV}^2$ for
atmospheric mixing \cite{Ashie:2005ik,Maltoni:2004ei} we find for
a regular hierarchy $m_1 \sim m_2 < 0.05\eV$ and $m_3<0.07\eV$, while
for 2-sigma lower limit $\Delta m^2_{13}=1.8\times 10^{-3}\eV^2$ one finds
$m_3<0.065\eV$. At 2-sigma the degeneracy is thus lifted at the 30\% level,
$m_3/m_1>1.3$.  For an inverted hierarchy one has $m_1 \sim m_2<0.062\eV$
and $m_3<0.045\eV$ for $\Delta m^2_{13}=1.8\times 10^{-3}\eV^2$ and again the
degeneracy between $m_1$ and $m_3$ is lifted at the 40\% level with
2-sigma confidence. This is illustrated in Figure \ref{fig:bongo}.

It is interesting to note that with another factor of 2 improvement we
will be able to distinguish a normal hierarchy (for which $\sum m_{\nu}
\sim 0.05\eV$) from an inverted hierarchy (with $\sum m_{\nu} \sim
0.1\eV$), even in the limit of one massless neutrino.  This is also
shown in Figure \ref{fig:bongo}, where the 68\% confidence limit is
already below the minimal inverted hierarchy solution and an inverted 
hierarchy is disfavored at 80\% confidence.  We
cannot yet differentiate between these two cases in the data with any
statistical significance. 
We also note that assuming one or the other neutrino mass makes very little
difference in the estimates of other data as there are no strong
degeneracies with other parameters present.

%% some difference in
%% the estimates of other parameters. This is best seen from Figure
%% \ref{fig:nn2d}, which shows 2-d contours between $\sum m_{\nu}$ and
%% remaining cosmological parameters.

%% \begin{figure}
%%   \centering
%%   \includegraphics[width=\linewidth]{neutn_2D.ps}
%%   \caption{This figure shows the masses $\sum m_\nu$ against remaining
%%   parameters. Do we need this??}
%%   \label{fig:nn2d}
%% \end{figure}

In addition to the strong evidence for mixing from solar and
atmospheric neutrino experiments there is a hint for another
oscillation with a much larger mass-squared difference coming from
the LSND experiment \cite{1996PhRvL..77.3082A}.  This requires a mass
squared difference of around $1\eV^2$, which cannot be incorporated in the
standard 3 neutrino scheme.  Of all the schemes that could incorporate
this result the one that is most plausible is 3+1, three active and
one sterile neutrino.  However, even this scenario is strongly
constrained by the short baseline oscillation experiments and the
window is nearly closed at 3-sigma level \cite{Maltoni:2004ei}.
Existing cosmological constraints suggest that a sterile neutrino with
mass at or above 0.8eV is ruled out at 2-sigma, but not yet at 3-sigma
\cite{2005astro.ph.11500D}.  So it is worth revisiting this issue with
the new analysis.

Our new analysis considerably strengthen these limits. We find $m_\nu
<0.26 \eV$ at 95\% c.l. (0.43eV at 99.9\%).  
This completely rules out a thermalized sterile neutrino with
mass around 1eV as the explanation for LSND experiment. However, it is
possible to suppress the production of sterile neutrinos in the early
universe, for example by large lepton asymmetry \cite{Foot:1996qc}.
Therefore, there exists a possibility to accommodate these results with LSND
by postulating that the neutrinos are not thermalized and their
concentration is significantly suppressed relative to a thermal
distribution.
   Note that even though the LSND-kind of sterile neutrino is disfavoured, the
  data in general prefer more than 3 standard neutrinos worth of
  relativistic species in the early universe. Thus a light sterile 
neutrino may well be preferred by the data, as 
  might be other light particles.

\subsection{Curvature}

Curvature is strongly constrained by CMB data because it determines
the angular position of the acoustic feature through its effect on the
angular diameter distance to the last scattering surface.  However,
other quantities such as the cosmological constant and matter to
radiation ratio also have a similar, albeit less pronounced effect, so
the CMB alone cannot determine these parameters separately with a high
accuracy. Adding additional information, such as the baryonic acoustic
horizon, Hubble constant or redshift-distance relation from
supernovae, improves the constraints considerably. In our analysis we
find
\begin{equation}
\Omega_{\rm k}=-0.003^{+0.0060}_{-0.0061}{}^{+0.0109}_{-0.0122}{}^{+0.0157}_{-0.0180}
\end{equation}
when curvature is added to the basic 6 parameters.  This is the
strongest constraint on curvature to date and the data continue to
show no evidence for it.  It should be noted that the curvature
constraints are affected by the choice of parameters and adding some,
like a time dependent equation of state, can weaken these limits
considerably.  On the other hand, inflation as the leading paradigm
for structure formation predicts no curvature, so there is
good justification to ignore it in the remaining analyses.

\subsection{Dark energy}

Dark energy affects the rate of growth of structure, especially
for $z<1$ where the dark energy is dynamically important.
We have assumed in the analysis so far that the dark energy equation 
of state is ${\rm w}=-1$. Here we relax this and explore 
the constraints on w. In this paper we only explore the simplest 
case of a constant equation of state. 
It has been shown in \cite{2005PhRvD..71j3515S}
that the constraints assuming a constant w are very similar 
to constraints obtained for time dependent 
w in a 3-parameter expansion of the equation of 
state as a function of expansion factor, assuming these are given 
at redshift 0.2-0.3, the pivot point for equation of state information.
So our constraint on w is more general in the
sense that it remains valid for smooth time dependencies of 
the equation of state, assuming it is 
interpreted as the value at z=0.2-0.3. 
We find 
\begin{equation}
{\rm w}=-1.040^{+0.063}_{-0.063}{}^{+0.124}_{-0.130}{}^{+0.178}_{-0.208} . 
\label{w}
\end{equation}
This is an improvement over other recent constraints and we see that
${\rm w}=-1$ remains very close to the best fit.  Note that we have
included perturbations in the dark energy both for ${\rm w}<-1$ and
${\rm w}>-1$, although this makes very little difference in the
constraints given how close the results are to ${\rm w}=-1$, where
perturbations vanish.  
Most of the improvement comes from the reduced
error in the amplitude of fluctuations provided by the Ly-$\alpha$ forest, 
which helps
when combined with other data sets.  Our results prefer somewhat more
negative values for w than the WMAP analysis \cite{2006astro.ph..3449S},
closer to the cosmological constant ${\rm w}=-1$.  As a result
those tracker quintessence models which
predict ${\rm w} \sim -0.7$ \cite{1999PhRvD..59l3504S} appear
to be strongly excluded.  Other dark energy models which predict ${\rm
  w}\sim -1$ remain acceptable.

It is interesting to investigate the sensitivity of the constraint to 
the inclusion or exclusion of supernovae data. 
Our results above include both supernovae from 
the SNLS sample \cite{2006A&A...447...31A} and from the 
gold sample of \cite{2004ApJ...607..665R}. 
Without the latter we find
${\rm w}=-1.05 \pm 0.07$, while without the former we find ${\rm w}=-1.01 \pm 0.07$ 
and the two are statistically compatible. 
Finally, without any supernovae data at all the constraint is ${\rm w}=-1.02\pm 0.12$, 
so the supernovae reduce the error by a factor of two relative to the other data. 
In all cases cosmological constant is the preferred solution lying 
within one statistical deviation of the best fit value. 

\subsection{Iso-curvature and cosmic strings}
\label{isocos}
Finally, we constrain parameters of two possible extensions to the
inflationary paradigm. Firstly we constrain the baryon or CDM
iso-curvature modes, assuming complete correlation (or anti-correlation)
and equal scale dependence of perturbations. 
Parameter $A_{\rm iso}$ is the ratio of amplitude
in these fully correlated baryon/CDM iso-curvature modes to that in the
adiabatic modes. We find
\begin{eqnarray}
  A_{\rm iso,bar}=  -0.06^{+0.18}_{-0.18}{}^{+0.35}_{-0.34}{}^{+0.50}_{-0.55} \\
  A_{\rm iso,CDM}= -0.007^{+0.034}_{-0.035}{}^{+0.068}_{-0.067}{}^{+0.110}_{-0.104}. 
\end{eqnarray}
This is about a factor of 2 better than previous limits \cite{Beltran:2005gr}. 
For models predicting baryon isocurvature perturbations correlated
with scalar perturbations from inflation the predictions for $A_{\rm
  iso,bar}$ are to be of order unity \cite{Boubekeur:2006nj}, which is
severely constrained by the data. Curvaton model induced $A_{\rm
  iso,CDM}$ give a larger range of possibilities, some of which remain
viable even with the new data and corresponding analyses \cite{Beltran:2005gr}.

We also constrain the amount of cosmic strings allowed by the
present data. We predict a cosmic string power spectrum using 
calculations in \cite{Pogosian:1999np,pogoerr,Seljak:2006hi}, for a set of
their fiducial cosmological and string network parameters. 
The only free parameter is the string tension $G\mu$
which accounts for a simple quadratic scaling in amplitude. We find
that the present data require
\begin{equation}
  G\mu<2.3\times 10^{-7} {\rm at}\,  95\% {\rm c.l.} \, (<2.9\times 10^{-7}\, {\rm at} \, 99.9\%).
\end{equation}
This is comparable to other recent limits \cite{pogoerr,Fraisse:2006xc}, but
is subject to several uncertainties in the string network evolution, which 
may affect the relation between string tension and CMB amplitude. 

\section{Discussion and Conclusions}

In this paper we combine the current Ly-$\alpha$ forest constraints from SDSS, 
supplemented by a small amount of high resolution data, with
the constraints from other cosmological tracers, such as galaxy
clustering, supernovae and CMB, including the recent 3 year WMAP
analysis. Our Ly-$\alpha$ forest analysis is an improvement over the analysis
presented in \cite{2005ApJ...635..761M}. 
The main change is the inclusion of the mean
flux constraints (presented in \cite{2005ApJ...635..761M} but not used in the
main power spectrum analysis) derived from a principal component analysis of
quasar spectra, which allows one to identify the continuum level of
quasars up to an overall amplitude by using the information from about
10,000 quasar spectra.  These constraints are in remarkable
agreement with those derived from the power spectrum analysis,
providing additional support for the consistency of the overall
picture of the Ly-$\alpha$ forest.  When the two are combined the resulting constraints
on the amplitude and slope improve by about 20-30\%.

When adding this Ly-$\alpha$ forest information to other probes we
find significant improvements on many of the parameters.  For the
running of the spectral index we find $\alpha = -0.015 \pm 0.012$,
compared to $\alpha = -0.06 \pm 0.03$ without the Ly-$\alpha$ forest. Our
new value is in a good agreement with the previous analysis prior to WMAP3
data \cite{2005PhRvD..71j3515S} and in both cases provides no support
for running of the spectral index being non-zero. 
Our results are also in agreement with a similar analysis in \cite{Viel:2006yh}. 

The most dramatic improvements are found for neutrino masses.  This is
expected, since the Ly-$\alpha$ forest is one of the few tracers of small
scale power that is able to determine the amplitude of fluctuations in
dark matter on small scales, where massive neutrinos suppress the growth
of structure.  For the first time the constraints are sufficiently
strong that, when combined with atmospheric and solar neutrinos, they
lift the degeneracy of neutrino masses at the 30\% level.  The current
constraints are beginning to distinguish 
between a normal and inverted hierarchy in the limit of zero
mass for the lowest mass neutrino, with the inverted hierarchy 
being disfavored at 80\% confidence. Finally, we see no evidence for a
fourth massive neutrino that would explain the LSND experiment.  The
upper limit on its mass is $0.26$eV at 95\% confidence limit, which
rules out the entire range of parameter space for the LSND experiment,
assuming these neutrinos were thermalized in the
early universe. On the other hand, a massless thermalized sterile neutrino 
is favored, since we find 
$N_\nu^{\rm eff}=5.3^{+0.4}_{-0.6}{}^{+2.1}_{-1.7}{}^{+3.8}_{-2.5}$. 

Adding Ly-$\alpha$ forest data to the mix favors the simplest
possible model without running, massive neutrinos, strings, 
isocurvature perturbations, curvature or dark
energy equation of state different from -1. The only exception is 
the number of relativistic degrees of freedom, where the data 
prefer additional contribution beyond photons and 3 neutrino families.  
In many cases LYA brings the value of the parameters 
closer to the expected value than the analysis without LYA.
As discussed in this
paper, there is some tension between the Ly-$\alpha$ forest and CMB in the
amplitude of fluctuations, but the statistical significance is small
and the two data sets are compatible in amplitude at 
$\sim 2$-sigma.  Any statistical assessment of compatibility must also take
into account the a-posteriori nature of the comparison, since the data
are measuring several parameters and one should not focus only on the
most discrepant one to assess compatibility.  It is worth mentioning
that the amplitude tension of the two data sets is at a similar level
of significance as the evidence for running or curvature in the WMAP
data, which many (especially inflation theorists) would be willing to
accept as a statistical fluctuation until proven otherwise.  This is
expected: one is estimating a dozen parameters and it is not
surprising if some are 2-sigma away from their true values.  In fact,
the evidence for running is eliminated after Ly-$\alpha$
forest data is added and the error is reduced by a factor of 
2-3.  
So one can either accept that the original
evidence was a statistical fluctuation which was settled conclusively
with Ly-$\alpha$ forest and other data, or one can argue that some of these
data have issues with systematic errors and should be excluded from
the analysis.  In this paper we have proceeded under the 
assumption that these are just normal statistical fluctuations and that the
data are compatible with each other.  Future observations and investigations
of possible systematic effects will provide a more definitive answer.

We wish to conclude with a brief discussion of the role of unaccounted
systematics in the current cosmological analyses.  No cosmological
probe is completely immune to systematic uncertainties and all must be
subject to ongoing investigations, as well as cross-checks among
different data sets.  The remarkable progress in statistical precision
of cosmological data over the past few years has been unprecedented,
but often the improved statistics of the data sets has not been
adequately matched by correspondingly improved analysis of
systematic errors. To some extent this is expected: systematic errors
are more difficult to track down and quantify and a detailed analysis
of these can take years, so is often done after the initial results
have been published. However, such analysis is necessary and provides
important confidence in the results. A lot of progress has been made
on this front in the past years and in some cases these concerns have
been lessened after a detailed study of systematic errors has been
published.  A case in point is the remarkably detailed analysis of
systematics in the recent WMAP 3 year data, which has 
strengthened the reliability of cosmological constraints
derived from WMAP data.  It is essential that corresponding
investigations are done with other tracers as well and we believe that
there is some room for improvement in many cases. For some
cosmological tracers, such as supernovae and weak lensing, the open
issues (e.g. intrinsic alignments and shear calibration in weak
lensing, calibration and evolution in SN) are known and well organized
campaigns to address them are underway.  For other tracers,
such as galaxy clustering, uncertainties in galaxy formation already
limit the statistical power that can be extracted on small scales, but
the full extent of this uncertainty needs to be quantified better.

Since the key results of this paper depend crucially on the
reliability of the Ly-$\alpha$ forest it is particularly important to
address the systematics of this tracer.  Recent investigations of
the Ly-$\alpha$ forest have shown that possible astrophysical
complications are either small or can be incorporated into the
analysis
\cite{2005MNRAS.360.1471M,2004MNRAS.350.1107M,2004ApJ...610..642C,Lai:2005ha}.
Moreover, an independent analysis of the SDSS flux power spectrum
obtained results that are in agreement with the original analysis
\cite{2006MNRAS.365..231V}, although a somewhat lower amplitude is 
favored \cite{Viel:2006yh} (note that one does not expect exactly the same
result because \cite{2006MNRAS.365..231V} used only
SDSS data, while we add high resolution data that helps break parameter 
degeneracies).  
However, there are remaining worries
about possible other effects that have not yet been thought of and
more investigations are needed to exhaust the list of possibilities.
%It is worth emphasizing that the current analysis prefers the simplest
%cosmological model with no running, no massive neutrinos, no
%curvature, no deviations of the dark energy equation of state away from
%${\rm w}=-1$, no cosmic strings or isocurvature modes and, as
%previously established, no evidence for warm dark matter
%\cite{Seljak:2006qw}.  In many cases the improvements due to
%Ly-$\alpha$ forest have been significant.  So if the actual universe
%has non vanishing value for one of these at a level that is strongly
%incompatible with these constraints then this means that the unknown
%systematic effect has to exactly cancel their effect to explain the
%results found here.  This is contrived, but not impossible and only
%with more exhaustive investigations we will gain understanding of
%whether this can happen or not.

One area where progress in Ly-$\alpha$ forest is expected soon is in
numerical simulations.  The original analysis in
\cite{2005ApJ...635..761M} used an indirect approach patching
together several simulations with differing box sizes to cover the
whole dynamical range in observations and relying on hydro-PM
simulations calibrated by only a few fully hydrodynamic simulations to
cover the relevant range of parameter space.  With the fast hydrodynamic
codes optimized for the Ly-$\alpha$ forest that are available today, such
as Enzo \citep{2004astro.ph..3044O} or the Eulerian moving frame
TVD+PM code described in \citep{2004NewA....9..443T}, this is no
longer necessary and parameter space can be covered
entirely with hydrodynamic simulations.  In addition, a code
comparison project focusing on Ly-$\alpha$ forest is underway and will
address the issue of agreement among different hydrodynamic codes.

We expect that as the progress in statistical precision of
cosmological constraints slows down in the future more emphasis will
be given to the investigation of systematic errors.  Until then the
current cosmological constraints obtained from combining many
different cosmological probes, including the ones presented in this
paper, should be viewed as preliminary. However, the fact that the
different tracers agree with each other 
within the range of normal statistical fluctuations
bodes well for the current round of
analyses and suggests systematic effects may be subdominant.

We thank Alexey Makarov for help with the MCMC code and we acknowledge
use of CAMB, CMBFAST and CosmoMC packages, as well as LAMBDA supported
by NASA office of Space Science.  Some computations were performed on CITA's 
Mckenzie cluster which was funded by the Canada Foundation for Innovation and 
the Ontario Innovation Trust \citep{2003astro.ph..5109D}.
U.S. is supported by the Packard
Foundation, NASA NAG5-1993 and NSF CAREER-0132953. AS is supported by
the Slovenian Research Agency grant Z1-6657.

\bibliography{cosmo,cosmo_preprints}
\end{document}